\documentclass{llncs}
\usepackage{graphicx}
\begin{document}

\title{Stabilizing and Destabilizing Effects of Embedding 3-node Subgraphs on the State Space of Boolean Networks}
\author{Chikoo Oosawa \inst{1,*}, Michael A. Savageau \inst{2}, Abdul S. Jarrah \inst{3}, \\Reinhard C. Laubenbacher \inst{3} and Eduardo D. Sontag \inst{4}}
\institute{Department of Bioscience and Bioinformatics, \\Kyushu Institute of Technology, Fukuoka, Japan\and Department of Biomedical Engineering, University of California, Davis,\\ California, U.S.A. \and Virginia Bioinformatics Institute, Department of Mathematics, \\Virginia Polytechnic Institute and State University, Virginia, U.S.A. \and Department of Mathematics, Rutgers, The State University of New Jersey, \\New Jersey, U.S.A.}
\maketitle
\subsection*{\centering Abstract}
We demonstrate the effects of embedding subgraphs in a Boolean network, which is one of the discrete dynamic models for transcriptional regulatory networks. After comparing the dynamic properties of networks embedded with seven different subgraphs including feedback and feedforward subgraphs, we found that complexity of the state space increases with longer lengths of attractors, and the number of attractors is reduced for networks with more feedforward subgraphs. In addition, feedforward subgraphs can provide higher mutual information with lower entropy in a temporal program of gene expression. Networks with the other six subgraphs show opposite effects on network dynamics. This is roughly consistent with Thomas's conjecture. These results suggest that feedforward subgraph is favorable local structure in complex biological networks.\\
\\Keywords : Boolean networks; subgraph; feedback; feedforward; mutual information; entropy; transcriptional regulatory networks; Thomas's conjecture

\section{Introduction}
Complex networks of interacting elements arising in biological, sociological, and physical areas can often be abstracted to graphs or networks. Recent studies of networks \cite{NETBIO,BOOK}, including transcriptional regulatory networks in cells, have revealed at least two statistical properties: power-law connectivity distributions having a small number of highly connected nodes; highly clustered connections among adjacent nodes \cite{Alon,Milo}. The last local structures, called subgraphs or motifs, consist of a few nodes and edges among the nodes that are statistically significant, and can be regarded as functional modules \cite{Milo}. Since feedback and feedforward subgraphs are basic and ubiquitous circuits in man-made systems, one can expect that transcriptional regulatory networks also have both feedback and feedforward subgraphs. However, only the feedforward subgraphs prevail \cite{Milo}. Other biological networks such as signal transduction and neuronal networks have similar tendencies. This suggests that feedforward subgraphs are favored in complex biological networks. In general, although the accumulated data of complex networks underlies the statistical significance, it is unclear why feedforward subgraphs are advantageous over other subgraphs in biological systems. We therefore constructed Boolean networks and embedded subgraphs in them to investigate their effects on both network structure and dynamics.\\
\section{Model and method}
\subsection{Boolean network}
The dynamics of the Boolean networks \cite{SAK93,PHD02} is determined by the equation
\begin{equation}
X_i(t+1)=B_i\left[{\bf X}(t)\right]\quad(i=1,2,...,N),
\label{eq:bn}
\end{equation}
where $X_{i}(t)$ is a binary state, either 0 or 1, of node {\it i} at time $t$, $B_i(\cdot)$ are Boolean functions [see Table \ref{tab:bool}] used to update the state of node {\it i}, and ${\bf X}(t)$ is a binary vector that gives the states of the $N$ nodes in the network. After assigning the initial states ${\bf X}(0)$ to the nodes, their successive states are updated by input states coming from upstream nodes and their Boolean functions. The dynamic behavior of these networks is represented by a time series of binary states. The time course follows a transient phase from its initial state until it establishes a periodic pattern, called an attractor [Fig. \ref{fig:state-space}].
\begin{table}[hbt]
\caption{16 Boolean functions with indegree $K_{in}$ = 2. In this paper, we used only No. 1, 2, 4, and 8 Boolean functions shown below, because of biological bias for the Boolean functions \cite{HARRIS02,LRAEY02,Alon03,NIKO07} and the feasibility of computation.}
\begin{center}
\begin{tabular}{|cc|*{16}{c}|}\hline
\multicolumn{2}{|c|}{Inputs}&\multicolumn{16}{c|}{Output}\\ \hline
0&0&0&0&0&0&0&0&0&0&1&1&1&1&1&1&1&1\\
0&1&0&0&0&0&1&1&1&1&0&0&0&0&1&1&1&1\\
1&0&0&0&1&1&0&0&1&1&0&0&1&1&0&0&1&1\\
1&1&0&1&0&1&0&1&0&1&0&1&0&1&0&1&0&1\\
\hline
Type&No.&\ 0&\ 1&\ 2&\ 3&\ 4&\ 5&\ 6&\ 7&\ 8&\ 9&10&11&12&13&14&15\\
\hline
\end{tabular}
\label{tab:bool}
\end{center}
\end{table}
\subsection{Numerical condition} % Sec 2.2 %
\begin{figure}[h]
\begin{center}
\includegraphics[scale=0.38]{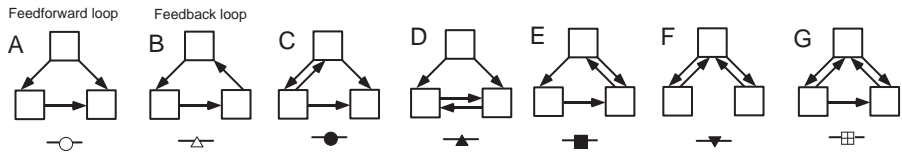} % Fig.1 %
\caption{Seven different subgraphs (local structures), The subgraphs consist of 3 nodes (squares) and more than 3 directed edges (arrows). Subgraphs A and B have three internal edges. Subgraphs C -- F have four and G has five internal edges. The number of cyclic loops increases with the number of internal edges. The seven subgraphs can be divided into two groups; subgraphs C, D, E, and G comprise subgraph A, feedforward loop; subgraphs B and F do not.}
\label{fig:7subs}
\end{center}
\end{figure}

To investigate the effects of embedding subgraphs on the Boolean network dynamics, we randomly constructed many networks with varying numbers of independent subgraphs. Seven subgraphs consisted of three nodes and more than three directed edges [Fig. \ref{fig:7subs} and Table \ref{tab:con}]. After embedding the specified number of subgraphs, the rest of the directed edges were assigned at random.
\begin{table}[h] % Table 2%
\begin{center}
\caption{Numerical condition: All constructed networks consist of the same amount of network resources, i.e., nodes, directed edges, and Boolean functions. Please note that the difference among the generated networks lies in the style of their connections.}
\label{tab:con}
\begin{tabular}{c|c}
\hline
Size of networks, $N$ & 128 nodes\\
Connectivity & For all nodes, $K_{in}$ = $K_{out}$ = 2\\
Boolean function & Only AND style [Table \ref{tab:bool}]\\
Types of subgraph & Seven different subgraphs [Fig. \ref{fig:7subs}]\\
Number of embedded subgraphs & 0, 10, 20, 30, and 40 \\
Number of edges & 256 \\
Number of initial states & 2000 per network\\
Number of realizations & More than $3\times 10^{3}$ in each condition\\
\hline
\end{tabular}
\end{center}
\end{table}
\subsection{Path length} % Sec 2.3 %
To obtain the structural changes of the propagating pathway of the state variables after embedding subgraphs, we measured the average path length \cite{AROB08}, which is the average number of the path lengths for all the nodes [Fig. \ref{fig:emb-pls}b], where the path length is the average number of directed edges in the shortest path from a node to all reachable nodes.
\subsection{Complexity of state space} % Sec 2.4 %
\begin{figure}[htb]
\begin{center}
\includegraphics[scale=0.40]{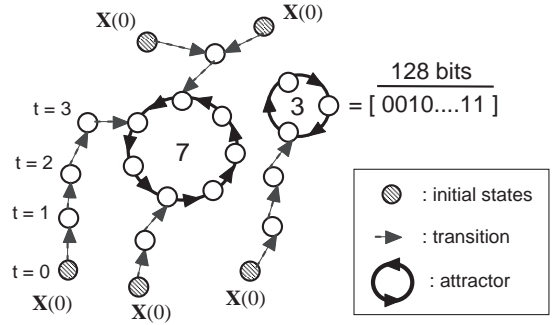} % Fig. 2 %
\caption{Example of a state space, There are $2^{128}(\sim10^{39})$ different states (shown as circles) in the space of each network. The numbers 3 and 7, inside the attractors, indicate the attractor lengths. We applied 2000 different initial states (shaded circles) to each network [Table \ref{tab:con}]. t = 0, 1,...., indicate time steps of Eq. (\ref{eq:bn}).} 
\label{fig:state-space}
\end{center}
\end{figure}

We use two measures to characterize the complexity of state space from the initial states [See Fig. \ref{fig:state-space}]: 
\begin{enumerate}
\item Basin entropy \cite{AROB08,ENTROPY}:
\begin{equation}
H_{Basin}=-\sum_{i} p(i)\log_{2}p(i)
\label{eq:be}
\end{equation}
where, the $p(i)$ satisfies $\sum_{i} p(i)=1$, $p(i)=\frac{a_{i}}{2000}$, and $a_{i}$ is the number of initial states that reached the i-th attractor [Fig. \ref{fig:state-space}].
\item Sum of the length of attractors: Each network may contain a different number of attractors, and their lengths may vary. This measure defines the total length of attractors in state space.
\end{enumerate}
The two measures indicate the complexity of state space from the initial states. According to the definitions, the larger values of two characteristics sgnify a higher complexity of the state space [Figs. \ref{fig:state-space} and \ref{fig:emb-atts}].
\subsection{Entropy and mutual information} % Sec 2.5 %
\quad We measured the entropy (randomness) and mutual information 
(correlation) of state variables to characterize the temporal structure of state variables in the Boolean networks 
\cite{PHD02,AROB07,NDES07}. Both dynamic properties are obtained from attractors [Figs. \ref{fig:state-space} and \ref{fig:emb-ens}].
\section{Results} % Sec 3 %
\begin{figure}[h]
\begin{center}
\includegraphics[scale=0.22]{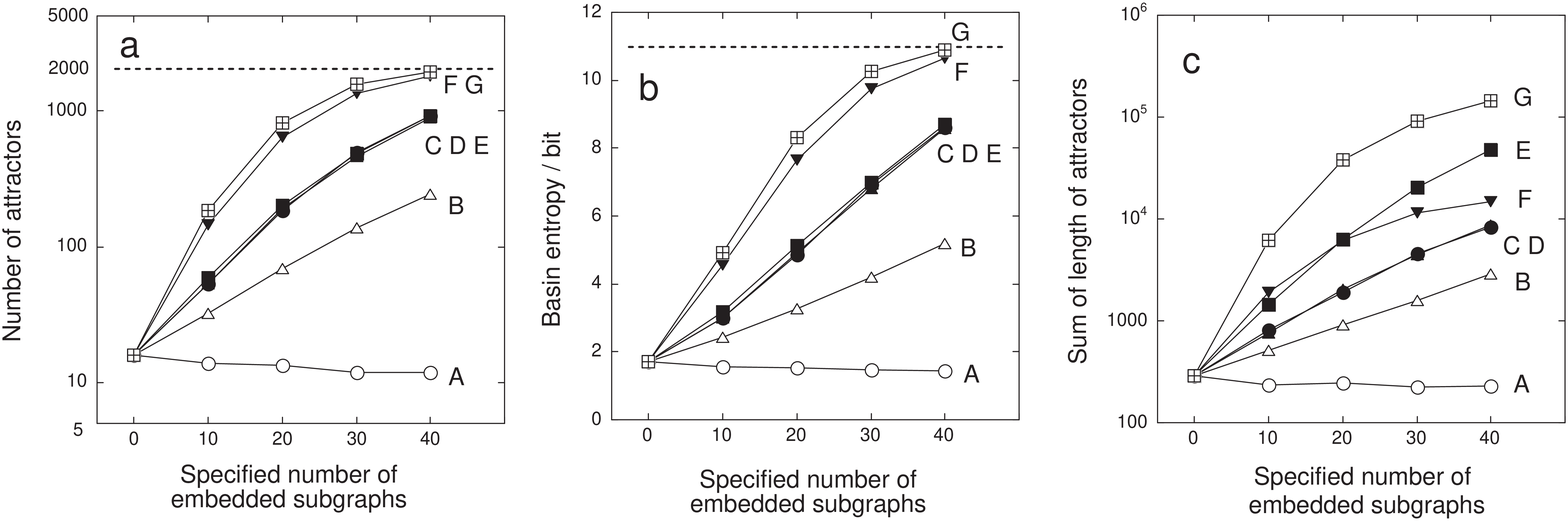} % Fig. 4 %
\caption{Complexity of state space structure. The relationship between the number of embedded subgraphs and the number of attractors (a), basin entropy (b) [See Eq.(\ref{eq:be})], and sum of the length of attractors (c). The symbols indicate mean values. Different capital alphabets indicate different subgraphs [Fig. \ref{fig:7subs}]. Dashed lines indicate the maximum value based on our condition [Table \ref{tab:con}]. The curves approaching the dashed lines indicate underestimates of the numbers of attractors.}
\label{fig:emb-atts}
\end{center}
\end{figure}

\begin{figure}[h]
\begin{center}
\includegraphics[scale=0.22]{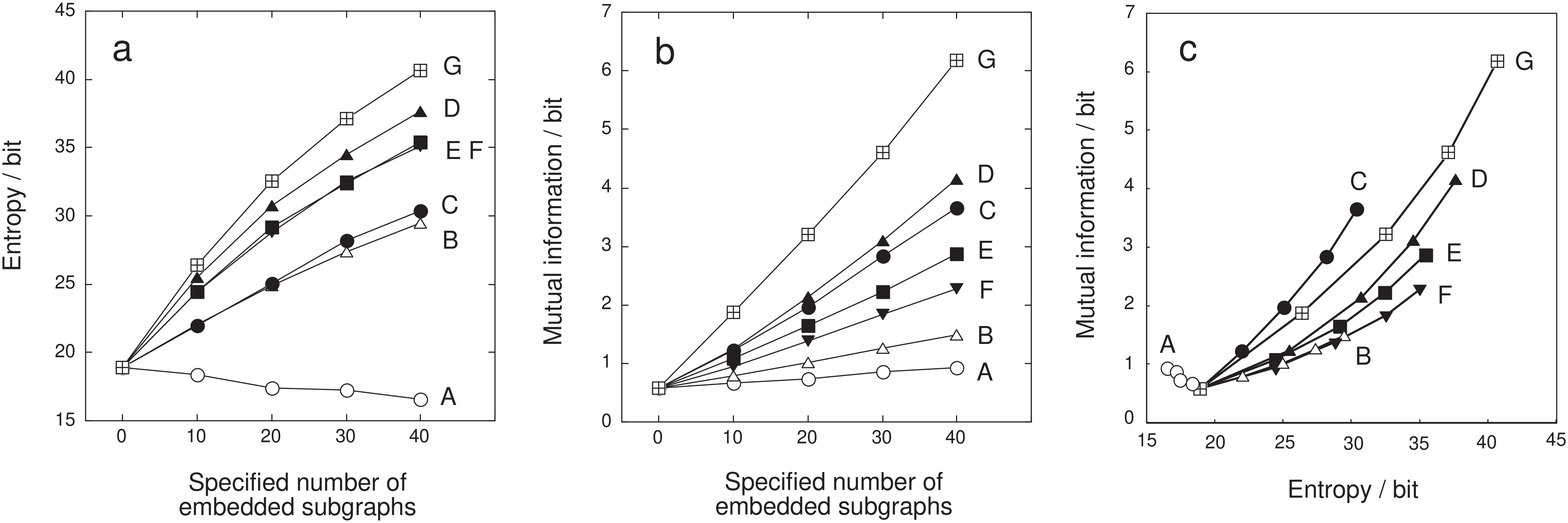}  % Fig. 5 %
\caption{Temporal structure of state variables. The relationship between the number of embedded subgraphs and the amount of entropy (a), the amount of mutual information (b), and rearranged data from a and b, (c). The symbols indicate mean values. Different capital alphabets indicate different subgraphs [Fig. \ref{fig:7subs}].}
\label{fig:emb-ens}
\end{center}
\end{figure}

\label{sec:nr}
\begin{figure}[h]
\begin{center}
\includegraphics[scale=0.25]{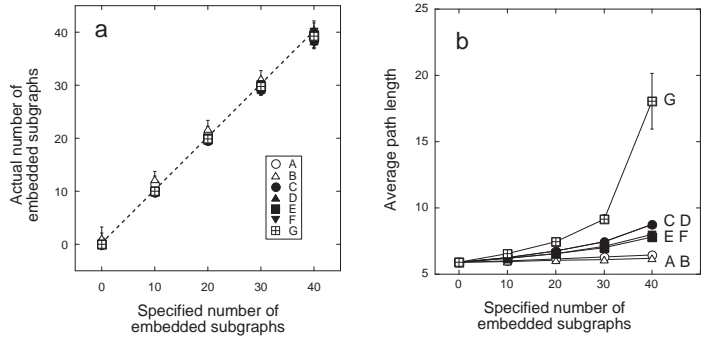} % Fig. 3 %
\caption{Structural properties of embedded networks. a: Relationship between the specified number of embedded subgraphs and the actual number of embedded subgraphs. Dashed line is given by $y = x$. b: Relationship between the number of embedded subgraphs and average path length. Symbols indicate mean and error bars show SD. Different capital alphabets indicate different subgraphs [See Fig. \ref{fig:7subs}].}
\label{fig:emb-pls}
\end{center}
\end{figure}
Since Fig. \ref{fig:emb-pls}a shows that our method was successful in embedding subgraphs in Boolean networks, the horizontal axis of Figs. \ref{fig:emb-atts}, \ref{fig:emb-ens}a, and \ref{fig:emb-ens}b are the specified number of embedded subgraphs. Because the number of inter-subgraph edges decrease as the number of internal edges increase, the average path length prolongs as the number of embedded subgraphs increases [Fig. \ref{fig:7subs} and \ref{fig:emb-pls}b].
%  Results %

We examined the effects of embedding subgraphs on the graphic and dynamic properties. Figure \ref{fig:emb-atts} shows the complexity of state space structures. Feedforward has opposite effects from the other six subgraphs. Note that the number of attractors, basin entropy, and the sum of the lengths of attractors decrease slightly with feedforward subgraphs. In general, the complexity of state space increases as the number of internal edges increase [See Fig. \ref{fig:7subs}]. The temporal structure of state variables is shown in Fig. \ref{fig:emb-ens}. The entropies in Fig. \ref{fig:emb-ens}a exhibit similar tendencies to those shown in Fig. \ref{fig:emb-atts}. Unlike Figs. \ref{fig:emb-atts} and \ref{fig:emb-ens}a, the amount of mutual information increases with an increase in the number of embedded subgraphs [Fig. \ref{fig:emb-ens}b]. Figure \ref{fig:emb-ens}c shows rearranged results from Fig. \ref{fig:emb-ens}a and \ref{fig:emb-ens}b, indicating the productivity of mutual information (correlation) from entropy (randomness).
\section{Discussion} % Sec 4 %
Based on resultant dynamics of networks as shown in Figs. \ref{fig:emb-atts} and \ref{fig:emb-ens}, the effects of embedding subgraphs in Boolean networks can be divided into two groups. Networks with more subgraphs, excluding feedforward, show a larger number of attractors and greater entropy and mutual information, demonstrating that the six subgraphs increase the complexity of the state space of the networks. In other words, the subgraphs behave as destabilizers of state space, pattern generators of a temporal program of gene expression, or entropy generators. The resultant mutual information (correlation) is driven by entropy as shown in the six positive slopes in Fig. \ref{fig:emb-ens}c. These results are consistent with the Thomas's and Sontag's conjecture \cite{CONJECTURE,Sontag1,Sontag2} because the reciprocal edges in a subgraph contribute to increasing of cyclic loops in the subgraph [Fig. \ref{fig:7subs}].

On the other hand, networks with more feedforward subgraphs show a smaller number of attractors and less entropy, but greater mutual information. This indicates that feedforward subgraphs stabilize the state space as well as organize temporal patterns with less entropy, as shown by the negative slope in Fig. \ref{fig:emb-ens}c. Detailed analyses with differential equations \cite{FFL1,FFL2} show feedforward loops are robust to variations in biochemical parameters and work as a low-pass filter. Together with our numerical results and related work \cite{AROB08,Sontag2}, this suggests that feedforward loops are favorable local structure in complex biological networks.

Actual complex biological networks are established based on emergence and evolutionary processes, and the resultant structure has many statistical features. Here we concentrate on the effects of embedding subgraphs in Boolean networks. The constructive approach \cite{PHD02,AROB08,AROB07,NDES07,Sontag2} promises to provide insight into the prediction of relationships between network structures, behaviors, and functions.
\subsection{Control parameters for Boolean networks} % Sec 4.1 %
The control parameters for the dynamics of Boolean networks are input connectivity, $K_{in}$, the size of network, the bias of the Boolean functions, and output connectivity distributions \cite{PHD02,AAM05,MAPC03}. In this report, we change only the connection style, while maintaining the same amount of network resources [Table \ref{tab:con}]. Figures \ref{fig:emb-atts} and \ref{fig:emb-ens} demonstrate that the internal connection style, such as the number of reciprocal edges or cyclic loops, may well be regarded as a novel control parameter for the dynamics of Boolean networks.
\subsection{Differences in correlation productivity} % Sec 4.2 %
In general, complex adaptive systems, including biological systems, perform their functions correctly when certain appropriate communications among nodes are established, because such systems need to add or delete nodes, or change the connectivity strength to adapt to exogeneous inputs optimally. Therefore, the productivity of correlation among nodes is the critical factor for the networks. However, the dynamics of our results are collective behavior of the interaction of single kind of subgraphs. The lowest correlation productivities in Fig. \ref{fig:emb-ens}c can be seen for subgraphs B and F, these two subgraphs do not involve feedforward structures [Fig. \ref{fig:7subs}]. The other subgraphs, excluding subgraph A, show greater productivity. In fact subgraph C and D have the largest freqencies in signal transduction, neuronal networks \cite{Alon,Milo}.

\subsection*{Acknowledgements}
This work was supported by a Grant-in-Aid for Young Scientists (B) No. 18740237 from MEXT, Japan (C.O.).

\end{document}